\documentstyle[epsfig,longtable]{aipproc}

\renewcommand{\[}{\begin{equation}}
\renewcommand{\]}{\end{equation}}

\begin{document}

\setlength{\unitlength}{1cm}
\noindent \begin{picture}(0,0)
\put(0,2){\noindent \sf Presented at the:}
\put(0,1.5){\noindent \sf ``Fifth Williamsburg Workshop on
First--Principles Calculations for Ferroelectrics'',}
\put(0,1){\noindent \sf February 1--4, Williamsburg, VA, USA}
\end{picture}
\title{Hartree--Fock Studies \\ of the Ferroelectric
Perovskites}

\author{L. Fu,$^1$ E. Yaschenko,$^1$ L. Resca,$^1$ and R.
Resta$^{1,2}$}
\address{$^1$Department of Physics, The Catholic University of
America, Washington, D.C. 20064 \\$^2$INFM--Dipartimento di
Fisica Teorica, Universit\`a di Trieste, Strada Costiera 11,
I--34014 Trieste, Italy}

\maketitle

\begin{abstract}  Within an {\it ab--initio} HF scheme, we use
both Berry--phase calculations and supercell calculations in
order to compute the dynamical charges for lattice dynamics and
the electronic dielectric constant for KNbO$_3$ and BaTiO$_3$. 
Comparison with experimental data indicates that HF
provides a description of the electronic properties of this
material whose accuracy is of the same order as the LDA one.
There are however significant differences between the two sets of
results, whose origin is scrutinized.  Motivated by the study of
surface and domain--boundary properties, we also present some
results for BaTiO$_3$ slabs, including both genuinely isolated and
periodically repeated slabs with different terminations.  The
capability of dealing with a genuinely isolated slab is a virtue
of the localized--basis implementation adopted here. We
demonstrate, amongst other things, the nontrivial dynamical--charge
neutrality of BaTiO$_3$ [001] surfaces.
\end{abstract}

\section*{Introduction}

Density--functional theory (DFT) has been very successful in
predicting several physical properties of the ferroelectric
perovskites, most implementations being performed within the
local--density approximation (LDA). Here we assess the
performance of the alternative Hartree--Fock (HF) scheme for
these materials, which have an intermediate ionic/covalent
character. In general, DFT is particularly convenient and
accurate for metals and semiconductors~\cite{Cohen91}, while the
HF scheme is very reliable for ionic crystals, and insulators in
general~\cite{Pisani,Pisani2}.  In this paper, we address at the
HF level some fundamental dielectric properties, such as the
dynamical ionic charges and the dielectric constant, in two
paradigmatic ferroelectric perovskites: KNbO$_3$ and BaTiO$_3$.

The very first HF study of a ferroelectric perovskite, at the {\it
ab--initio} level, has been published very recently~\cite{rap97}. 
The case study was KNbO$_3$, and several properties of the
electronic ground state were investigated, including the
broken-symmetry instability of the tetragonal structure of this
material. In this same work the Berry--phase theory of macroscopic
polarization\cite{modern,rap_a12} is implemented within the HF
scheme, using first-principles ingredients.  Perhaps the most
interesting result that has emerged is that the calculated HF
spontaneous polarization of KNbO$_3$ in its tetragonal phase is
0.34 C/m$^2$, quite close to the experimental value\cite{expt},
which is 0.37 C/m$^2$, as well as previous LDA
results\cite{rap75,Zhong94,Wang96}, which are between 0.33 and
0.40 C/m$^2$.

Although the computed value of the spontaneous polarization is
almost identical within LDA and HF, we will show that there are
significant differences when the dynamical charges of the ions are
separately investigated.  These polarizabilities are dominated by
covalency effects, which plausibly are described in different ways
at the HF and LDA levels.

We will then show that, by combining results from both
Berry--phase calculations and supercell calculations, one has
access to the theoretical value of the electronic dielectric
constant $\varepsilon_\infty$~\cite{rap99}. It is well known that
LDA overestimates the value of $\varepsilon_\infty$ in any
material, whereas it has been only guessed that HF underestimates
it, since no calculation was previously available, for any
material. Our HF calculated $\varepsilon_\infty$ of KNbO$_3$,
BaTiO$_3$ and MgO are indeed smaller than experimental values.

Finally, we will show some results for BaTiO$_3$ slabs, including
both isolated and periodic slabs with different terminations. 
Since most first--principles schemes are implemented within
plane--wave based computational methods, they are very effective
for infinite periodic systems but cannot treat the isolated slabs.
Even the periodic slabs with moderate vacuum separations prove to
be a heavy computational burden for plane--wave methods, and only
a few such calculations have been performed  for ferroelectric
perovskites\cite{Cohen97,Padilla97}.  The capability of dealing
with a genuinely isolated slab is a virtue of the localized--basis
implementation adopted here~\cite{Pisani,crys}. We will discuss
bulk dielectric properties, surface charge, surface energies, work
function, and surface dynamical charges.  We will show how the
computed values of these quantities change as functions of the
number of atomic layers and the thickness of the vacuum layers
separating the periodic slabs.  Owing to a recently discovered
theorem, the dynamical charge of a given ion at a polar surface
{\it cannot} have the same value as in the bulk~\cite{rap94}.

\section*{Bulk dynamical charges and dielectric constant}

The transverse (Born) charge tensor is defined via the macroscopic
polarization $\Delta {\bf P}$ linearly induced by a rigid
displacement ${\bf u}_s$ of the $s$ sublattice, while the field is
kept at zero value\cite{PCM}. Namely, \begin{eqnarray} \Delta
P_{\alpha} ={1\over{\Omega}}\sum_{\beta} Z_{\alpha \beta}^{\ast
({\rm T})}u_{\beta}, \end{eqnarray} where $\Omega$ is the cell
volume, and atomic units are used throughout.  The longitudinal
(Callen) charge tensor $Z^{\ast ({\rm L})}$ is similarly defined,
but with the sublattice displacement performed in a depolarization
field $\Delta {\bf E}= -4\pi \Delta {\bf P}$. The relation between
the two dynamical charges is \begin{eqnarray} Z^{\ast ({\rm
T})}=\epsilon_{\infty}Z^{\ast ({\rm L})}, \label{lt} \end{eqnarray}
where $\epsilon_{\infty}$ is the dielectric constant in the
relevant direction. In general, the dynamical charge can be very
different from the nominal static charge of an ion. The differences
are particularly dramatic when the material has a mixed
ionic--covalent character, such as the materials chosen in the
present work~\cite{rap84,rap90}.

We calculate the $Z^{\ast ({\rm T})}$ as a
Berry--phase\cite{modern,rap_a12}, while we calculate the $Z^{\ast
({\rm L})}$ using a supercell technique. The latter approach was
introduced by Martin and Kunc several years ago to compute dynamic
charges in semiconductors\cite{Kunc}, and is still very much in use
today in the framework of LDA pseudopotential calculations with
plane--wave basis sets. The same supercell technique in the
framework of HF calculations with a localized basis set was
implemented for the first time in Ref.  \cite{rap99}. Since the aim
of the present work is to assess the HF values versus LDA ones, we
only present here results for the cubic structure. For KNbO$_3$ we
use the lattice constant 4.016 \AA.  This is the same value used in
the LDA study of Ref.~\cite{Wang96}, which is based on the
low--temperature experimental data\cite{expt}. For BaTiO$_3$, we
use the lattice constant 4.006 \AA .  Figure 1 shows the structure
of these perovskites.

\begin{figure}[t] \begin{center} \setlength{\unitlength}{1cm}
\begin{picture}(14.9,5.5) \put(0,0){\includegraphics{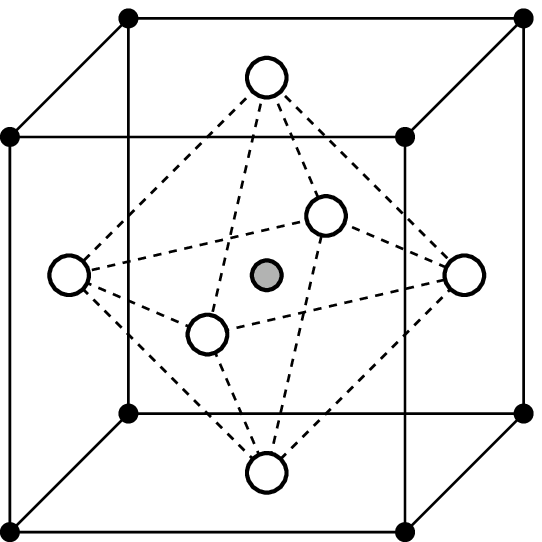}}
\put(7,5.5){\parbox[t]{7.8cm}{Figure 1. Cubic perovskite structure,
with general formula ABO$_3$, where in our cases A is either K or
Ba (solid circles) and B is either Nb or Ti (shaded circle). The
oxygens (empty circles) form octahedral cages, with B at their
centers, and arranged in a simple cubic pattern. We will consider
either displacements in the vertical direction, or slabs with
horizontal surfaces. In both cases ``O$_{{\rm I}}$'' refers to top
and bottom oxygen ions and ``O$_{{\rm II}}$'' refers to the four
oxygen ions in the basal plane of the octahedron. We have
therefore AO$_{{\rm I}}$ and BO$_{{\rm II}}$ planes.  }}
\end{picture}\end{center} \end{figure}

We use the {\sc Crystal95} code, which implements both the HF and
LDA self consistent scheme with a
linear--combination--of-atomic--orbitals (LCAO) basis
set\cite{crys}. In our supercell calculations, we displace the
sublattice of an atom in the supercell lattice, and compute the
planar average of the charge density, potential, and field. We
then perform additionally the macroscopic average defined as in
Ref.~\cite{macro}, which proves an invaluable tool in the present
business. The output of the supercell calculations provide the
longitudinal dynamical charges.  We use both 3--fold and 4--fold
supercells in our supercell calculations, which yield almost
identical results. We refer to the original paper for more
details~\cite{rap99}. 

In our Berry--phase calculations, we compute the induced
macroscopic polarization $\Delta {\bf P}$ in zero field, and thus
the corresponding transverse dynamical charges. In order to compare
with the LDA results and examine the performance of our localized
basis set, we also run the {\sc Crystal95} code in the LDA option
for KNbO$_3$. Our results for KNbO$_3$ dynamic charges are reported
in Table I.  \begin{center} Table I. Dynamical charges for
KNbO$_3$ \\ \smallskip \begin{tabular}{||l|c|c|c|c|c||} \hline
&K&Nb&O$_{\rm I}$&O$_{\rm II}$&sum\\ \hline Z$^{\ast ({\rm L})}$,
HF, 3--fold supercell &0.377&2.766&$-$1.848&$-$0.624&0.047\\ \hline
Z$^{\ast ({\rm L})}$, LDA, 3--fold supercell
&0.187&1.696&$-$1.227&$-$0.335&$-$0.014\\ \hline Z$^{\ast ({\rm
T})}$, HF, Berry--phase &0.830&8.702&$-$6.130&$-$1.703&$-$0.004\\
\hline Z$^{\ast ({\rm T})}$, LDA, Berry--phase
&1.144&10.400&$-$8.377&$-$1.567&0.033\\ \hline Z$^{\ast ({\rm
T})}$, LDA, Ref.~\protect\cite{Wang96}
&1.12&9.67&$-$7.28&$-$1.74&0.03\\ \hline \end{tabular} \end{center}
\smallskip The last column shows that the acoustic sum
rule\cite{PCM} is satisfactorily obeyed.  A few calculations for
$Z^{\ast ({\rm T})}$ of KNbO$_3$ at LDA level
exist\cite{rap75,Zhong94,Wang96}: these data are slightly scattered
and in some cases refer to slightly different geometries.  Here we
compare our results with Ref. \cite{Wang96}, which uses the same
geometry. Our LDA $Z^{\ast ({\rm T})}$ are in reasonable agreement
with those of Ref. \cite{Wang96}. The small differences may be due
to the fact that our LDA exchange-correlation functional differs
somewhat from that of Ref.  \cite{Wang96}, and our pseudopotentials
and basis sets are optimized for HF rather than LDA.  

To determine the dielectric constant $\epsilon_{\infty}$, we plot
$|Z^{\ast ({\rm T})}|$ versus $|Z^{\ast ({\rm L})}|$, and use a
least--square fitting to determine the slope which yields
$\epsilon_{\infty}$. This is shown in  Fig. 2, which indicates
that the correlation between our $Z^{\ast ({\rm L})}$ and $Z^{\ast
({\rm T})}$ is reasonably linear.  For KNbO$_3$ we obtain the LDA
value $\epsilon_{\infty}=6.33$, which is close to the LDA value of
6.63 obtained in Ref. \cite{Wang96}, and is about 35 \% higher
than the experimental value 4.69. Similarly, we obtain the HF
value $\epsilon_{\infty}=3.17$, which is about 32 \% smaller than
the experimental value.  

\begin{figure}[t] \begin{center} \setlength{\unitlength}{1cm}
\begin{picture}(14.9,4.3) \put(0,-1){\includegraphics{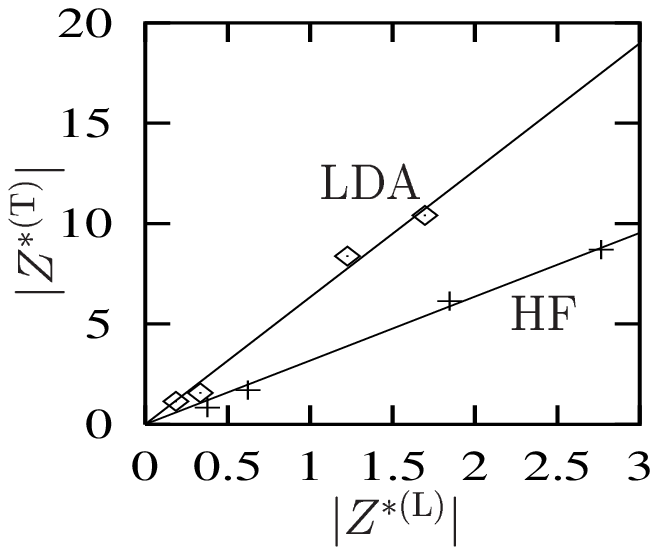}}
\put(7.5,3.3){\parbox[t]{7.3cm}{Figure 2. The transverse charge
$|Z^{\ast ({\rm T})}|$ versus longitudinal charge $|Z^{\ast ({\rm
L})}|$. The slopes of the two solid lines are, respectively, the
LDA and HF dielectric constant $\epsilon_{\infty}$.}}
\end{picture} \end{center} \end{figure}

\begin{center} Table II. Dynamical charges for BaTiO$_3$\\
\smallskip \begin{tabular}{||l|c|c|c|c|c||} \hline &Ba&Ti&O$_{\rm
I}$&O$_{\rm II}$&sum\\ \hline Z$^{\ast ({\rm L})}$, HF, 3--fold
supercell &0.909&2.126&$-$1.468&$-$0.797&$-$0.027\\ \hline Z$^{\ast
({\rm T})}$, HF, Berry--phase
&2.052&5.996&$-$4.319&$-$1.846&$-$0.037\\ \hline Z$^{\ast ({\rm
T})}$, LDA, Ref.~\protect\cite{Ghosez}
&2.74&7.29&$-$5.75&$-$2.13&0.02\\ \hline \end{tabular} \end{center}
\smallskip We have performed analogous calculations for BaTiO$_3$,
whose results are reported in Table II: the resulting HF dielectric
constant is $\epsilon_{\infty}=2.76$. This value is about one half
of the experimental value 5.40. The LDA value reported in
Ref.~\cite{Ghosez}, for a slightly different lattice constant, is
6.73.  We have also performed a series of similar calculations for
the more simple oxide MgO\cite{rap101}. We use the cubic lattice
constant $a=4.21$ \AA. We obtain HF $Z^{\ast ({\rm L})}=\pm 1.267$
and $Z^{\ast ({\rm T})}=\pm 1.809$, which yields a HF
$\epsilon_{\infty}=1.428$. This value is about one half of the
experimental value 2.94, while the LDA value reported in
Ref.~\cite{Schutt94} is 3.14.

As a general conclusion about the study of bulk dielectric
properties, we find that the calculated transverse dynamical
charges are about of the same quality within HF and LDA, the trend
being that the HF ones are generally closer to the nominal value of
the corresponding static charges. This seems to indicate that HF
tends to underestimate the covalence mechanism, while LDA tends to
overestimate it. The resuts for the dielectric constants confirm
this same message.
 
\section*{Surface properties via slab calculations}

\begin{figure}[b] \begin{center} \setlength{\unitlength}{1cm}
\begin{picture}(14.9,3.4) \put(0,-2){\includegraphics{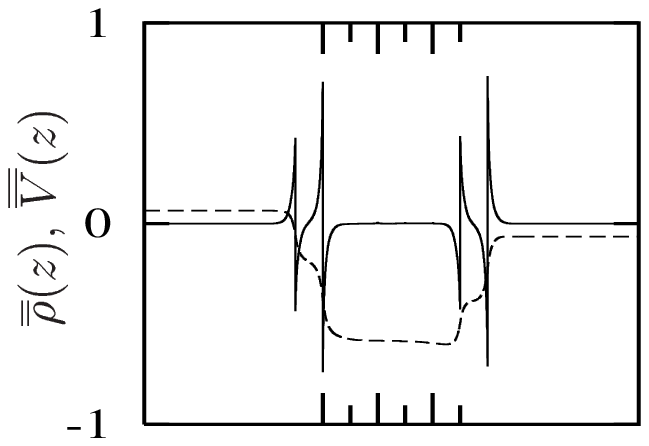}}
\put(7.5,3.4){\parbox[t]{7.3cm}{Figure 3. The macroscopic average
of charge density (10$^{-1}$ electron/bohr$^3$, solid line) and
potential (hartree, dashed line) for an isolated
slab of six atomic layers, where the bulk structure is
centrosymmetric. The vertical bars indicate the positions of the
atomic layers: short bars for BaO$_{\rm I}$ planes, long bars for
TiO$_{\rm II}$ planes.  }} \end{picture} \end{center} \end{figure} 

We now switch to the HF investigation of several surface
properties of BaTiO$_3$. To this aim we have used both isolated
slabs and periodically repeated ones, while at this preliminary
stage surface geometries have not been relaxed: all results refer
therefore to ideal (truncated bulk) geometries. There are two
major differences between the two kinds of calculations.  First of
all, there are interactions among the periodic slabs, which
presumably vanish when the vacuum separating the slabs is
sufficiently thick. Using a basis of plane waves (or
plane-wave-like, such as LAPW), the increase in the vacuum space
drastically increase the number of plane waves necessary for
convergence, and therefore the computational workload. Secondly,
the isolated slabs are calculated by imposing the boundary
condition of vanishing external field, while the periodic slabs
are necessarily calculated with the boundary condition of
vanishing {\it average} field (over the supercell). This is a
major problem when dealing with supercells built of asymmetrically
terminated slabs, even in absence of macroscopic polarization. The
reason is that the work functions of the two surfaces are
generally different. The supercell is an artificial periodic
structure of repeated slabs and vacuum layers, where the
electrostatic potential is enforced to be periodical. This
periodicity, combined with the vacuum--level difference (due to
the different work functions), necessarily originates fictitious
electric fields and {\it static} surface charges. Such unphysical
charges can in principle be made arbitrarily small by increasing
the thickness of the vacuum layers, but in practice a large enough
supercell may be computationally unaffordable.  

\begin{figure}[t] \begin{center} \setlength{\unitlength}{1cm}
\begin{picture}(14.9,3.4) \put(0,-2){\includegraphics{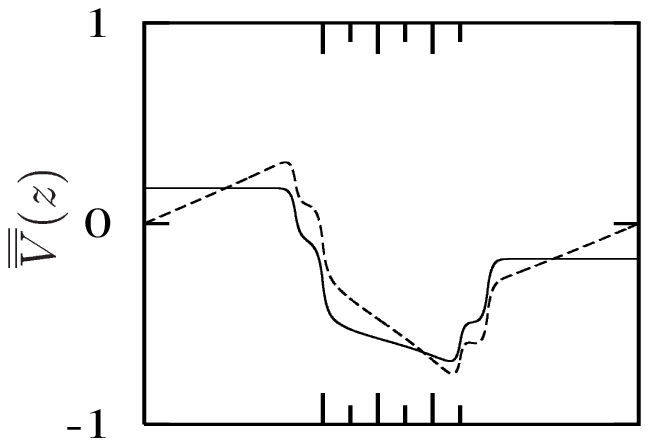}}
\put(7.5,3.2){\parbox[t]{7.3cm}{Figure 4. Macroscopic averages of
the potential (hartree) for an isolated slab (solid
line) and a periodically repeated slab (dashed line).  Both
calculations refer to slabs cut from a noncentrosymmetric bulk
having the experimental ferroelectric distortions.  The meaning of
vertical bars is the same as in Fig. 3. }} \end{picture}
\end{center} \end{figure} 

The problem does not occur when performing calculations for
isolated slabs within a localized basis set. As an example we plot
in Fig. 3 the macroscopic average of charge density, potential,
and field for an isolated slab with asymmetric termination, where
however the slab has been cut from a centrosymmetric bulk. There
are several very perspicuous features to remark.  (1) First of
all, despite the asymmetric termination of the slab of Fig. 3, the
macroscopic field vanishes both in the bulk region and in the
vacuum region. This means that each of the two surfaces is
neutral, as required by very general considerations.  (2) One
observes some structure only in the surface regions.  Basically,
there is a quadruple charge layer (two double layers) at each
surface.  We also see that in the middle region of the
slab both the charge and field vanish and the potential remains
constant. This means that the bulk like behavior is recovered in
the middle region of the slab.  (3) Although the field in the
vacuum regions is zero, the potential goes to two different limits
to the left and to the right of the sample. This net change in the
potential across the slab is precisely the difference in the work
functions of the two nonequivalent surfaces, discussed above. (4)
We have additionally calculated all kinds of slabs (asymmetrically
terminated as in Fig. 3, doubly Ti-terminated, doubly
Ba-terminated): all calculated surface properties 
look precisely the same, while instead they artificially
{\it do} depend on the choice of the slab in the periodically
repeated case.

As a matter of fact, periodically repeated slabs are systematically
messed up by the above problems. We show here an example in Fig. 4,
where the occurrence of the fictitious field is very perspicuous,
despite the large thickness of the vacuum region (6 lattice
constants in this calculation).  Because of this major problem,
very few calculations of a periodically repeated slab
(asymmetrically terminated) have ever been performed, for any
material. For a discussion of the difficulties, see
Refs.~\cite{Cohen97,rap94}. By contrast with this, the
isolated--slab calculation conveys a very clear message, and the
slope of the potential in the bulk region of Fig. 4 is no artifact:
the presence of a macroscopic field is a physical feature, related
to the spontaneous polarization of the bulk material. This will be
discussed in the following.

\subsection*{Polarization, surface charge, and surface dipoles}

Let us consider the total dipole per unit area of a finite slab of
$n$ atomic layers: this dipole is easily evaluated from
isolated--slab calculations simply looking at the net change in
potential across the slab $\Delta V_{\rm slab}$, and is obviously
nonzero even for centrosymmetric slabs (with asymmetric
termination), as in Fig. 3. In the case of a ferroelectric slab, an
additional contribution to the dipole comes from the spontaneous
polarization. In the limit of large $n$, one must recover the
appropriate bulk polarization. Since the nominal thickness of the
slab is $na/2$, we write the dipole per unit volume as:
\begin{equation} P_n= \frac{\Delta V_{\rm slab}}{2\pi na} \simeq
\sigma_0+{{2p_0}\over{n}}. \end{equation} In the latter expression
$\sigma_0$ is the charge per unit area which accumulates at the
slab surface, and is a bulk property dependent on the spontaneous
polarization of the material. Instead $p_0$ is proportional to the
work function difference discussed above and is therefore a surface
property. Our results are displayed in Fig. 5, which shows the
asymptotic behavior of $P_n$: the surface charge $\sigma_0$ is
clearly zero in the centrosymmetric case, while we estimate
$P_\infty = \sigma_0 = 0.089 {\rm C/m}^2$ for the ferroelectric
isolated slab.

Besides the above asymptotic extrapolation, an alternative path to
evaluating $\sigma_0$ consists in measuring the macroscoic
electric field in the middle of the isolated slab. In the
centrosymmetric case we get zero field at any thickness, while the
value of $\sigma_0$ extracted in this way from the ferroelectric
slab calculation, shown in Fig. 5 as a triangle, is $\sigma_0 =
0.095 {\rm C/m}^2$, in good agreement with the above mentioned
value.

\begin{figure}[t] \begin{center} \setlength{\unitlength}{1cm}
\begin{picture}(14.9,4.2) \put(0,-1){\includegraphics{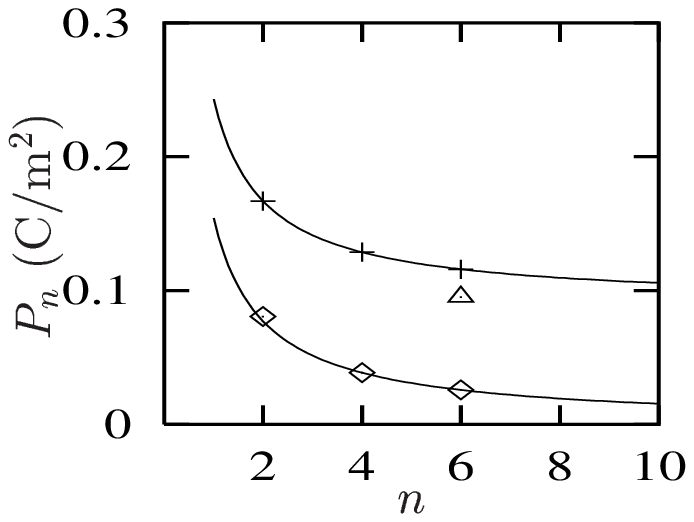}}
\put(7.5,3.8){\parbox[t]{7.3cm}{Figure 5.  Dipole moment per unit
volume as functions of the number of atomic layers, for isolated
centrosymmetric (diamonds) and ferroelectric (crosses) slabs. The
solid lines are fits of Eq. (4). The triangle is the bulk
polarization calculated from the field in the bulk region of the
ferroelectric slab.}} \end{picture} \end{center} \end{figure} 

Finally, we wish to compare the above calculated value of
$P_\infty$ to the bulk spontaneous polarization of the material, as
measured in experiments and as previously calculated within
LDA\cite{Zhong94,Ghosez}. It is important to notice that our
$P_\infty$ is by definition the macroscopic polarization in a
depolarizing field, while the literature generally refers to the
spontaneous polarization in zero field. The latter is related to
the former by a factor $\varepsilon_\infty$, much in the same way
as the longitudinal and transverse dynamical charges in
Eq.~(\ref{lt}).

Using our theoretical value of 2.76 for the HF dielectric constant
we get a spontaneous polarization of $0.245 {\rm C/m}^2$, which
compares extremely well to the experimental value of 
$0.263 {\rm C/m}^2$ \cite{Wieder55}, and to LDA values in the range
0.286 to 0.363 ${\rm C/m}^2$ \cite{Zhong94,Ghosez}.

In order to achieve a more meaningful comparison, we have also run
our own Berry phase calculation using identical technical
ingredients and geometry as in the slab calculations. We get a
value of $0.240 {\rm C/m}^2$, in very good agreement with the above
evaluation of the same quantity from the electric field inside the
isolated slab. This shows the internal consistency of the HF
properties, calculated at the HF level.  Indeed, the small
disagreement provides an estimate of the numerical error, within
the same given physical approximation. We emphasize that, on the
contrary, any periodic slab calculation would be affected by much
larger numerical errors.

\subsection*{Surface energy}

\begin{figure}[b] \begin{center} \setlength{\unitlength}{1cm}
\begin{picture}(14.9,4.4) \put(0,-1){\includegraphics{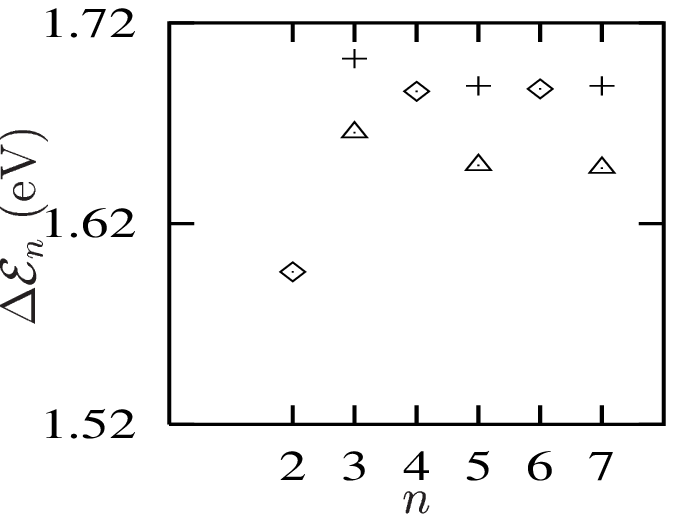}}
\put(7.5,3.9){\parbox[t]{7.3cm}{Figure 6.  Calculated surface
energy as functions of the number of atomic layers.  The diamonds
and crosses are, respectively, the results from isolated slabs
with asymmetric and symmetric terminations. The triangles are the
results from periodic slabs with symmetric terminations. }}
\end{picture} \end{center} \end{figure} 

We report here a study of the surface energies. We emphasize
that---since the surface geometries have not been relaxed---our
preliminary calculated values have not to be taken as realistic.
What we demonstrate instead is that using a genuinely isolated
slab we have a fast convergence to the genuine surface properties,
whereas convergence is rather slow in the periodic--slab geometry.
Our slabs, made of $n$ atomic layers cut from a centrosymmetric
bulk, have symmetric termination for odd $n$, and asymmetric
termination for even $n$.

In the present [001] geometry two nonequivalent truncations of the
bulk crystal are obviously possible. Whenever this happens---as
thoroughly discussed in the literature~\cite{Qian88}---the surface
energy of each of the nonequivalent surfaces is ill defined and
any stability issue must be studied by introducing the appropriate
chemical potential. We therefore address the average surface
energy (over the two nonequivalent surfaces).

The average surface energy $\Delta {\cal E}$ per surface unit cell
can be evaluated from slab calculations (with $n$ atomic layers)
as the asymptotic value of \begin{eqnarray} \Delta {\cal
E}_n=\left\{{ \begin{array}{ll} \left[{2{\cal E}_n-n{\cal E}_{\rm
bulk}}\right]/4,\ \ \ & \mbox{ even } n,\\ \left[{{\cal E}_n({\rm
Ti})+{\cal E}_n({\rm Ba}) -n{\cal E}_{\rm bulk}}\right]/4,\ \ \ &
\mbox{ odd } n,\\ \end{array}}\right. ,  \end{eqnarray} where ${\cal
E}_n({\rm Ti})$ and ${\cal E}_n({\rm Ba})$ indicate the energy of
a symmetrically terminated slab with Ti or Ba termination,
respectively.

From Fig. 6 we see that for isolated slabs, as $n$ increases, the
average surface energy converges very fast to the asymptotic value:
from above for odd $n$, and from below for even $n$. On the other
hand, the results for periodic slabs are sistematically lower than
those of isolated slabs. This indicates that interactions among
slabs still exist, even with a large vacuum separation: in all
calculations reported here, we use a fixed supercell size of 12
interlayer spacings. We mention that the surface energies obtained
in previous calculations at LDA level using periodic slabs are
0.921 eV \cite{Cohen97}) and 1.237 eV \cite{Padilla97} for a
relaxed geometry \cite{Cohen97}), while 1.358 eV is reported for
the unrelaxed geometry\cite{Padilla97}.  As stated above, our
geometry is unrelaxed: the data in Fig. 6 extrapolate to 1.69 eV.

\subsection*{Surface dynamical charges}

We evaluate the surface longitudinal dynamical charges Z$^{\ast
({\rm L})}$ of BaTiO$_3$ from HF calculations performed on
symmetrically--terminated slabs of seven atomic layers. The bulk
is centrosymmetric, and the reference surface geometry is not
relaxed.  Upon displacing atoms in the surface region, we get the
results in an analogous way as for the bulk longitudinal charges
discussed above. The dynamical charges of ions in the first and
second outermost surface layer are reported in Tab.  III, together
with the corresponding bulk values (taken from Tab.  II), for the
sake of comparison.

\begin{center} Table III. Surface $Z^{\ast ({\rm L})}$ (HF)\\
\smallskip \begin{tabular}{||c|c|c|c|c|c|c||} \hline
&Ba&Ti&O$_{\rm I}$&O$_{\rm II}$& Ba+O$_{\rm I}$& Ti+2O$_{\rm II}$
\\ \hline first layer
&1.104&2.093&$-$1.354&$-$0.919&$-$0.250&0.255\\ \hline second
layer &0.911&2.119&$-$1.473&$-$0.767&$-$0.562&0.585 \\ \hline bulk
&0.883&2.187&$-$1.543&$-$0.765&$-$0.660&0.657\\ \hline
\end{tabular} \end{center} We notice that BaO$_{\rm I}$ planes and
TiO$_{\rm II}$ planes are {\it nominally} charge neutral, in
the hypothesis of complete ionicity. But dynamical
charges---particularly in a material having mixed ionic--covalent
character as the present one---are dramatically different from the
static nominal ones. As a matter of fact, the data in the last
two columns of Tab. III
show that BaO$_{\rm I}$ planes and TiO$_{\rm II}$ planes are
far from being dynamically neutral, both in the bulk and in the
surface region.

As observed above, the surface of a centrosymmetric slab is
charge--neutral as far as the {\it static} charge accumulated in
the surface region is concerned. This fact implies---owing to a
recently discovered theorem~\cite{rap94}---that the surface region
must also be {\it dynamically} neutral. We notice that the
dynamical charges in the surface region deviate considerably from
their bulk value, although they converge rather fast to it. A
corollary of the dynamical--neutrality theorem, as applied to the
[001] geometry considered here, states the the sum of the
dynamical charges in the surface planes must add up to {\it one
half} of the corresponding bulk value (with the appropriate
sign)~\cite{rap94}. This sum rule---which can be regarded as the
surface analogue of the popular acoustic sum rule for the
bulk~\cite{PCM}---is very accurately fullfilled by our
calculations. In fact summing over the two outermost surface
layers we get $-0.250\!+\!0.585\!=\!0.335$ and
$0.255\!-\!0.562\!=\!-0.307$ for the two nonequivalent surfaces. 
Both figures are indeed very close (in modulus) to one half of the
dynamical charge of a given bulk plane.

\section*{Conclusions}

We have reported several {\it ab--initio} HF results concerning
bulk and surface properties of two important ferroelectric
perovskites: KNbO$_3$ and BaTiO$_3$. Though some of our results
are preliminary, they indeed assess the accuracy of both the
physical approximations and the computational implementation.
 
About the physical approximations, comparison of our results with
the experiment indicates an overall trend: the HF results are
about of the same quality as---but on the opposite side of---the
LDA results. This trend makes HF an appealing method for
``bracketing'' a theoretical prediction: it is well exemplified by
the case study of the dielectric constant $\varepsilon_\infty$
discussed here for KNbO$_3$, BaTiO$_3$, and even for the simpler
oxide MgO.

About the computational implementation within a localized basis
set, we have emphasized its merits by showing some case studies of
surface properties, where we can study genuinely isolated slabs in
vacuo. This is at variance with basically all previous
first--principle investigation of surface properties which, for any
material, were typically performed within supercell ({\it i.e.}
periodically repeated slab) geometries.

\section*{Acknowledgments}

We are indebted to  S. Dall'Olio and R. Dovesi for discussions
and help in using {\sc Crystal}95, as well as for providing us
with optimized values of the bulk computational parameters.  This
work is supported by the Office of Naval Research,
through grant N00014-96-1-0689.

\end{document}